\documentclass[aps,prl,twocolumn,groupedaddress,amsmath,amssymb,showpacs]{revtex4}

\usepackage{color}

\usepackage{graphicx}
\usepackage{hyperref}
\usepackage{pifont}

\begin{document}
\title{Efficient spin injection and giant magnetoresistance in Fe/MoS$_2$/Fe junctions}
\author{Kapildeb Dolui}
\thanks{These authors contributed equally to this work.}
\author{Awadhesh Narayan\footnotemark[1]}
\author{Ivan Rungger\footnotemark[1]}
\author{Stefano Sanvito}
\affiliation{School of Physics, AMBER and CRANN, Trinity College, Dublin 2, Ireland}
\date{\today}

\begin{abstract}
We demonstrate giant magnetoresistance in Fe/MoS$_2$/Fe junctions by means of \textit{ab-initio} transport calculations. 
We show that junctions incorporating either a mono- or a bi-layer of MoS$_2$ are metallic and that Fe acts as an efficient spin 
injector into MoS$_2$ with an efficiency of about 45\%. This is the result of the strong coupling between the Fe and S atoms
at the interface. For junctions of greater thickness a maximum magnetoresistance of $\sim$300\% is obtained, which 
remains robust with the applied bias as long as transport is in the tunneling limit. A general recipe for improving the 
magnetoresistance in spin valves incorporating layered transition metal dichalcogenides is proposed.
\end{abstract}

\maketitle

\textit{Introduction}. 
Layered transition metal dichalcogenides (TMDs) have proved to be a fertile ground for fundamental phenomena in 
solid-state physics, ranging from superconductivity to charge density waves to Mott transitions, as well as very promising 
for technological applications such as energy storage, catalysis, logic circuits and high performance electronic devices 
like field effect transistors~\cite{geim-2d,strano-review,chhowalla-review,kis-fet}. For instance in MoS$_2$, a prototypical 
layered TMD, the bandgap changes from indirect to direct when the thickness reduces from bulk to the mono-layer limit~\cite{heinz-gap}. 
Consequently, there emerges photoluminescence, and a potential for optoelectronic devices~\cite{wang-photo}. 
At the same time, due to presence of heavy elements and the lack of inversion symmetry, a large spin-orbit splitting appears in thin films with odd number of 
MoS$_2$ layers~\cite{yao-valley}. This distinct feature induces the coupling of spin and valley degrees of freedom in the valence and conduction
bands of monolayer MoS$_2$~\cite{valley1,valley2,valley3}.

To date most of the research on the electronic transport in TMDs has focused on lateral configurations, where  
electron motion is far beyond the ballistic regime. An exception is the fabrication of a vertical tunneling transistor incorporating
a-few-layered MoS$_2$ ribbon~\cite{ponomarenko-vertical,duan-vertical}, but no theoretical analysis has been associated to 
such experimental study. An intriguing prospect for TMDs-based vertical transport devices is that of fabricating ultra-thin
magnetic tunnel junctions (MTJs)~\cite{butler-mgo,sanvito-mgo}. These may offer the opportunity of realizing low-resistance
high-magnetoresistance devices, i.e. they may become an intriguing materials platform for several spintronics applications
both in the magnetic recording and the sensing arena. Furthermore, understanding spin-injection from transition metals to TMDs
is a crucial step for realizing the vision of spintronics on a \emph{flatland beyond graphene}, a vision recently energized by 
the characterization of Schottky barriers at the interface between MoS$_2$ and ferromagnetic metals~\cite{kawakami-ferro,dash-ferro}. The feasibility of using related two-dimensional materials, graphene and boron nitride (BN) as spacers in magnetic junctions has been studied theoretically~\cite{kelly,yazyev}. 
However, the existence of a large number of layered TMDs, which range from metals to semiconductors, to magnets 
and superconductors, offers a greater potential for tunability and engineering of the magnetoresistive properties. 
In contrast, graphene or BN leave little scope for future improvements, since it has proven very difficult to open 
a significant gap in graphene and to modulate the gap in BN. Therefore, moving from graphene and BN to TMDs 
opens an entire new avenue of possible material stacks with different combined properties.

\begin{figure}[bh]
\center
\includegraphics[width=8.0cm,clip=true]{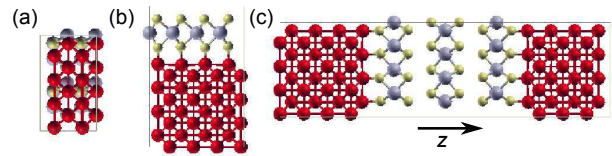}
\caption[]{(Color online) (a) A top view of the junctions investigated in this work, (b) side view of Fe/MoS$_2$ junction with single 
layer MoS$_2$ on Fe(001) substrate. (c) A side view of Fe/MoS$_2$-Fe junction for three layers of MoS$_2$ as spacer. We 
use a rotated supercell of MoS$_2$ in a rectangular geometry, with transport along $z$. The semi-infinite leads consist of Fe 
oriented along the (001) direction. Here red spheres denote Fe atoms, yellow spheres represent S atoms and blue spheres show Mo atoms.}
\label{fig1}
\end{figure}
\begin{figure*}[t]
\center
\includegraphics[width=16.0cm,clip=true]{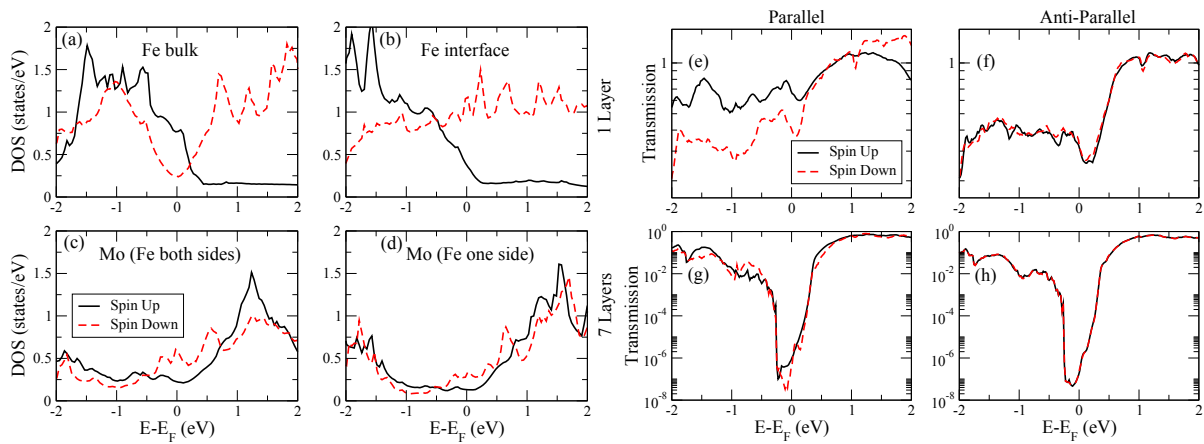}
\caption[]{(Color online) Densities of states projected on (a) the Fe atoms far away from the interface, (b) the Fe atoms at the interface, 
(c) the Mo atoms closest to the interface for a Fe/MoS$_2$/Fe junction and (d) the Mo atoms closest to the interface for a Fe/MoS$_2$
junction. Spin-resolved transmission as a function of energy for (e, f) one layer MoS$_2$ in both the parallel and anti-parallel configuration 
and (g, h) seven layers MoS$_2$ in both the parallel and anti-parallel configuration. Note that the up spin transmission at Fermi level is 
higher than that for the down spin for both one and seven layers of MoS$_2$ in the parallel configuration. For the anti-parallel case 
the transmission for the two spin channels is nearly identical.}
\label{fig2}
\end{figure*}

In this Rapid Communication we report our analysis on the transport across a few layer Fe/MoS$_2$/Fe MTJ devices. By means of first-principles
calculations we have discovered a giant magnetoresistance (MR) effect in Fe/MoS$_2$/Fe junctions, with a maximum MR 
of $\sim$300 \%, which remains robust with applied bias. Our calculations reveal that thinner junctions (spacers with mono- and 
a bi-layer MoS$_2$) are almost metallic, as a result of the strong coupling between MoS$_2$ and Fe surface. Importantly Fe electrodes efficiently inject carriers into MoS$_2$. Finally, we formulate a general recipe for obtaining higher MR, either by substituting the electrodes or alternatively by replacing the MoS$_2$ spacer with other layered TMDs.       


\textit{Computational Methods}. 
Our first-principles transport calculations are performed using the {\sc smeagol} 
code~\cite{sanvito-smeagol1, sanvito-smeagol2, sanvito-smeagol3}, which integrates the non-equilibrium Green's function method 
for electron transport with density functional theory~\cite{soler-siesta}. The core electrons are described by using norm-conserving 
pseudopotentials and we expand the electron density and the operators over a double-$\zeta$ polarized basis set. The real space 
mesh cutoff is 300~Ry and we consider the local density approximation (LDA) to the exchange-correlation functional. A rotated 
MoS$_2$ supercell is constructed in a rectangular geometry as shown in Fig.~\ref{fig1} and it is contacted by two semi-infinite 
(001)-oriented Fe leads. In our setup Fe is subject to a small strain ($\approx 4\%$) to make it commensurable with MoS$_2$, 
however this does not affect its electronic structure significantly. Growing Fe epitaxially on MoS$_2$ would realize this situation, 
while if MoS$_2$ is grown epitaxially on Fe, it will rather be strained to match the Fe lattice. If MoS$_2$ is transferred onto Fe after being 
prepared, for instance by exfoliation, we also expect it to retain its unstrained lattice constant. All geometries are relaxed until the forces 
on atoms are less than 0.02~eV/\AA. Periodic boundary conditions are employed in the plane perpendicular to the transport 
direction ($z$ direction), with a $12\times 8$ $k$-point grid for the self-consistent calculation. Transmission and densities of states are 
then obtained by integrating over a denser $120\times 80$ $k$-point mesh.

\textit{Results and discussion}. 
We begin our analysis by studying the density of states, DOS, projected on Fe away from the interface [see Fig.~\ref{fig2}(a)], which is very similar to unstrained case. The band gap variation of monolayer MoS$_2$ with strain is not expected to significantly alter the results, since the metallicity of the adsorbed MoS$_2$ layer is mainly due to the strong hybridization with the Fe substrate. For thicker junctions we expect the substrate-induced strain to affect mainly the interface layers, while for the subsequent MoS$_2$ layers it is progressively reduced, also because the interlayer coupling is determined by weak van der Waals forces. We note that if for a given metal/TMD interface the first TMD layer remains insulating, then strain provides a convenient means to engineer the TMD gap and with it the transport properties. The DOS of the interface Fe atoms has the usual peak in the down spin channel at the Fermi level, $E_\mathrm{F}$, however this is broadened due to 
the hybridization with MoS$_2$. The main contribution to the MoS$_2$ DOS around $E_\mathrm{F}$ originates from the Mo atoms. 
Such Mo DOS is comparable to the DOS associated to the Fe atoms as shown in Fig.~\ref{fig2}(c), and the native band gap of 
MoS$_2$ disappears in monolayer junction. Also, a strong bonding between MoS$_2$ and Fe (distance between S and Fe atoms is $\approx 1.9$ \AA{}), allows strong wavefunction overlap between Fe and Mo states, similar to that found for Ti and Mo contacts~\cite{tomanek-ti,banerjee-mo}. We note that for a lateral transport setup a significant Schottky barrier is formed when n-type MoS$_2$ is contacted by metal electrodes, which is directly related to the metal workfunction~\cite{das-schottky}. These findings are however specific of the lateral transport setup, with n-type MoS$_2$, and are not directly comparable to our results for pristine MoS$_2$ layers vertically intercalated between Fe. Moreover, the DOS at $E_\mathrm{F}$ becomes spin polarized, revealing spin injection into MoS$_2$, with an efficiency of $\eta=(\mathrm{DOS}_{\downarrow}-\mathrm{DOS}_{\uparrow})/(\mathrm{DOS}_{\uparrow}+\mathrm{DOS}_{\downarrow})\sim$45\% at the 
Fermi level. A similar situation is seen for the Fe/MoS$_2$ junction as well, with a similar figure for Fermi level 
spin injection efficiency. This suggests that a spin-polarized current can be injected even in a lateral transport setup. 

Next we calculate spin-resolved transmission as a function of energy for parallel and anti-parallel configurations of the
electrodes (the two magnetization vectors of the electrodes are either parallel or anti-parallel to each other). For a monolayer 
MoS$_2$ junction the transmission is high (conductance is of the order of one quantum $e^{2}/h$), indicating metallic 
transport as shown in Fig.~\ref{fig2}. At $E_\mathrm{F}$, the up spin transmission is higher than the down spin 
one in the parallel configuration, showing the role of Fe as an efficient majority spin injector for thin MoS$_2$ devices. In 
the anti-parallel configuration transmission in the two spin channels is nearly identical, with a small difference arising 
from the lack of inversion symmetry in our transport setup. We note that for a free-standing single MoS$_2$ layer LDA somewhat underestimates the quasi-particle gap when compared to GW calculations~\cite{gwbse,sanvito-substrate}. When the MoS$_2$ layer is deposited on Fe, the metal substrate provides large screening, which reduces the quasi-particle gap, so that it can be expected to get closer to the LDA value. Moreover, the strong hybridization with the Fe electrodes leads the first MoS$_2$ layer to be metallic, an effect rather insensitive to the exact gap of pristine MoS$_2$ itself. In contrast, for a bulk-like junction comprising of seven layers of MoS$_2$ the band gap of around 0.62~eV emerges, consistent with previous LDA calculations~\cite{sanvito-mos2ribbon}. Similar to the single layer case, at $E_\mathrm{F}$ the up spin transmission is greater than the down spin one in the parallel configuration. We have also performed calculations including spin-orbit coupling but our results remain essentially unchanged. 

\begin{figure}[ht]
\center
\includegraphics[width=8.0cm,clip=true]{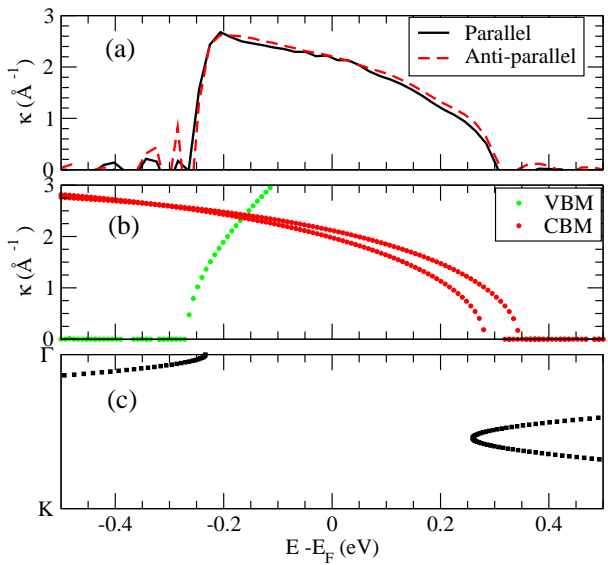}
\caption[]{(Color online) (a) The decay coefficient, $\kappa$, for the parallel and anti-parallel configuration calculated from the 
transmission function. Also shown are the bulk MoS$_2$ complex bands for the valence and the conduction band (b) along with 
the real band structure for comparison (c).}
\label{fig3}
\end{figure}

The wavefunction decay coefficient across the MoS$_2$ spacer, $\kappa_{ij}=\frac{1}{d_{i}-d_{j}}\ln \left( \frac{T_{j}}{T_{i}}\right)$, is 
calculated from the transmission coefficients for junctions with different thicknesses. Here $d_{i}$ is the thickness of a junction comprising $i$ layers, while $T_{i}$ is the corresponding transmission coefficient. As the spacer thickness increases the values of $\kappa_{ij}$ converge to a single bulk-like value, $\kappa$, for all used values of $i$ and $j$. In Fig.~\ref{fig3}(a) we plot this $\kappa$ for thick junctions ($i=9$ and $j=7$). Such decay coefficient
matches quite closely the evanescent wave-number obtained from the complex band structure of bulk MoS$_2$, which is plotted 
in Fig.~\ref{fig3}(b). The complex bands joining the real valence band maximum (VBM) and conduction band minimum (CBM) 
have quite distinct slopes and hence different effective masses. While the complex band associated with VBM has a larger 
slope, the one connecting to the CBM rises more gradually and in fact at $E_\mathrm{F}$, has a smaller decay coefficient of the 
two. Hence, at $E_\mathrm{F}$, transport in thicker junctions is conduction band dominated. 

\begin{figure}[hb]
\center
\includegraphics[width=8.0cm,clip=true]{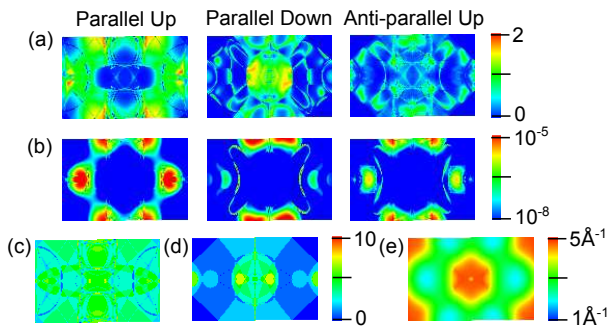}
\caption[]{(Color online) (a) $k$-resolved and spin-resolved transmission coefficient at the Fermi level for (a) one layer and 
(b) seven layers MoS$_2$. The left and central panels correspond to the parallel configuration while the right panel is for the 
anti-parallel set up. The $k$-resolved channels in the Fe leads at $E_{F}$ for up (c) and down (d) spins are plotted for 
comparison. The minimum value of $\kappa$ at the Fermi energy for bulk MoS$_2$ across the entire Brillouin zone is plotted
in (e).}
\label{fig4}
\end{figure}

To discern the $k$-resolved contributions to the transmission, in Fig.~\ref{fig4} we plot the transmission function at 
$E_\mathrm{F}$ across the 2D transverse Brillouin zone (BZ). The spin resolved transmissions are shown in panel (a) for 
the single layer MoS$_2$ junction. In the parallel configuration the up spin channel transmission mainly originates through 
the hexagonal region away from the BZ center, while for the down spin the transmission is mainly through a region close to the $\Gamma$ 
point. In comparison, for the anti-parallel case it is a combination of the previous two contributions. In anti-parallel 
configuration the down spin transmission is nearly identical to the up spin one and is not shown. Overall the transmission 
in both cases is fairly large, which opens up the possibility for spin injection into MoS$_2$. For seven layers junction 
the up spin transmission through the CBM ``hotspots'' is more pronounced as shown in Fig.~\ref{fig4}(b) and there is a marked 
reduction of the contributions from other $k$-points in the BZ. For the down spin and for the anti-parallel configuration the transmission at these points is greatly reduced. At the center of the band gap, $\kappa$ is smallest for the evanescent states connecting to the conduction band (Fig.~\ref{fig3}(b)). On the other hand, if $E_\mathrm{F}$ lies close to the valence band, a $\Gamma$ point dominated transmission is obtained from evanescent states originating from the valence band, opening a possibility to engineer preferential transmission from certain points in the BZ by modifying level alignment, for instance by doping.

We plot available channels in the Fe leads for up and down spin in Fig.~\ref{fig4}(c) and \ref{fig4}(d), respectively. The minimum $\kappa$ for the complex bands is shown in Fig.~\ref{fig4}(e) (smaller $\kappa$ implies larger transmission). This clearly reveals the six CBM ``hotspots'' apart from a smaller contribution for BZ center. In this tunneling limit the transmission through the junction depends both on the available states from the Fe leads as well as the decay of states in the MoS$_2$ spacer. Since for the parallel configuration these overlap in $k$-space over the BZ, the transmission is larger across such six CBM positions and the overall transmission is higher compared to the anti-parallel case. 

\begin{figure}[htb]
\center
\includegraphics[width=8.0cm,clip=true]{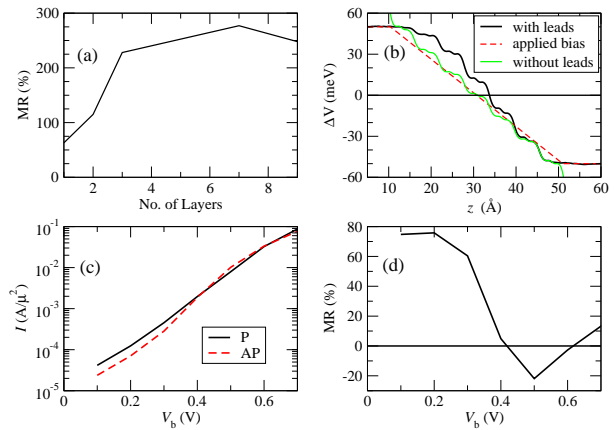}
\caption[]{(Color online) (a) Magnetoresistance as a function of the thickness of the MoS$_2$ spacer layer. Note that the MR 
saturates to a value of $\sim$300\% for nine layers. (b) The potential drop across the spacer showing the screening at the first 
MoS$_2$ layer in contact with Fe. The potential for a free MoS$_2$ seven-layer-slab under the same bias is also shown for 
comparison. The dashed red curve is the applied bias profile. (c) The current versus voltage plot for seven layer device. 
(d) The bias dependence of MR for seven layer spacer, showing its robustness, as long as one remains in the tunneling limit. }
\label{fig5}
\end{figure}

Based on our analysis of the specific case of Fe/MoS$_2$/Fe devices, we are now in the position to 
formulate a recipe for increasing the MR. For a larger MR using a MoS$_2$ spacer, one needs an electrode 
material with one spin predominantly found at the center of the BZ for energies around $E_\mathrm{F}$, while 
the other spin should be found preferentially at the six pockets away from $\Gamma$. This would selectively 
suppress conductivity through the spin channel at the center of the BZ, while allowing transmission for the 
other [see Fig.~\ref{fig4}(b)], therefore leading to a large spin filtering. Another strategy might be that of replacing 
MoS$_2$ with other two-dimensional TMDs. For large MR these should have a large band gap, minimum 
defects and smallest $\kappa$ in an area of the BZ where in the electrodes only one spin is dominantly 
present in the DOS at energies around $E_\mathrm{F}$. For single layer junctions we expect symmetry 
filtering at the interface to play an important role. In MoS$_2$ we find that at the $\Gamma$ point, $T$ is 
close to zero for majority spins in the parallel configuration, while it is rather large for the minority ones [Fig.~\ref{fig4}(a)], 
even though the bulk Fe electrodes provide states for both spins around the BZ center. This clearly shows 
that depending on the symmetry of the states in the electrodes the injection into the spacer changes dramatically.

The MR obtained for different MoS$_2$ spacers is summarized in Fig.~\ref{fig5}(a). For thinner junctions, MR is of the order of 100\%, which grows to $\sim$300\% on increasing spacer thickness, and saturates for the thickest junctions studied in this work. Usually MR results are presented only at zero bias, as in the previous two studies on graphene and BN~\cite{kelly,yazyev}. However, in a real device the bias dependence of MR is a key question, which we next turn our attention to. Fig.~\ref{fig5}(b) shows the potential drop across seven layer junction both with and without the Fe leads. In the presence of Fe leads the potential drop occurs only across the inner MoS$_2$ layers and remains flat for the first and last one. This is due to the screening by almost metallic layers in closest proximity to Fe, which is also an indication of ohmic nature of Fe/MoS$_2$ contact. The current versus voltage curve for the seven layer device is shown in Fig.~\ref{fig5}(c) for the parallel and anti-parallel configurations. For a range of bias up to 0.4~V, the current for the parallel configuration is higher than that for the anti-parallel one. This demonstrates the robustness of MR as long as one remains in the tunneling limit. Beyond such bias value one hits the conduction band edge and MR starts to decrease in an oscillatory fashion as the bias further increases. We note that LDA is known to underestimate the band gap of bulk MoS$_2$ by about a factor two, and thus the MR should, in principle, be robust for bias voltages larger than those predicted here.

\textit{Conclusion and outlook}. 
In conclusion, using fully-atomistic first-principles transport calculations, we have discovered a giant magnetoresistance effect 
in Fe/MoS$_2$/Fe junctions. We have found that Fe offers an efficient spin injection possibility for thinner junctions, which are 
metallic owing to a strong hybridization between Fe and interface S atoms. In thicker junctions the native gap of MoS$_2$ re-emerges 
and a robust MR is achieved as long as the transport remains in the tunneling limit. We have also formulated a general recipe to search for larger magnetoresistance in other layered materials. In addition, one may choose non-magnetic leads, but a magnetic spacer, like VS$_2$ or NbS$_2$~\cite{ciraci-tmd} to explore other possibilities. We are confident that our work will provide a guide to future studies, along the experimental front to fabricate our proposed device, as well as to theoretical investigations in a search for higher magnetoresistance based on two-dimensional layered materials.

\textit{Acknowledgments}. This work is supported by Science Foundation of Ireland (AMBER center) and by the Irish Research 
Council (AN). IR acknowledges financial support from the King Abdullah University of Science and Technology (ACRAB project). 
We thank Trinity Centre for High Performance Computing (TCHPC) and Irish Centre for High-End 
Computing (ICHEC) for providing the computational resources.



\begin{thebibliography}{100}

\bibitem{geim-2d} K.S.~Novoselov, D.~Jiang, F.~Schedin, T.J.~Booth, V.V.~Khotkevitch, S.V.~Morozov, and A.K.~Geim, Proc. Natl. Acad. Sci. U.S.A. {\bf 102}, 10451 (2005).

\bibitem{strano-review} Q.H.~Wang, K.~Kalantar-Zadeh, A.~Kis, J.N.~Coleman, and M.~Strano, Nat. Nano. {\bf 7}, 699 (2012).

\bibitem{chhowalla-review} M.~Chhowalla, H.S.~Shin, G.~Eda, L.-J.~Li, K.P.~Loh, and H.~Zhang, Nat. Chem. {\bf 5}, 263 (2013).

\bibitem{kis-fet} B.~Radisavlejevic, A.~Radenovic, J.~Brivio, V.~Giacometti, and A.~Kis, Nat. Nano. {\bf 6}, 147 (2011).

\bibitem{heinz-gap} K.F.~Mak, C.~Lee, J.~Hone, J.~Shan, and T.F.~Heinz, Phys. Rev. Lett. {\bf 105}, 136805 (2010).

\bibitem{wang-photo} A.~Splendiani, L.~Sun, Y.~Zhang, T.~Li, J.~Kim, C.-Y.~Chim, G.~Galli, and F.~Wang, Nano Lett. {\bf 10}, 1271 (2010).

\bibitem{yao-valley} D.~Xiao, G.-B.~Liu, W.~Feng, X.~Xu, and W.~Yao, Phys. Rev. Lett. {\bf 108}, 196802 (2012).

\bibitem{valley1} G.-B.~Liu, W.-Y.~Shan, Y.~Yao, and D.~Xiao, Phys. Rev. B {\bf 88}, 085433 (2013).

\bibitem{valley2} A.~Kormanyos, V.~Zolyomi, N.D.~Drummond, and G.~Burkard, Phys. Rev. X {\bf 4}, 011034 (2014).

\bibitem{valley3} K.~Kosmider, J.W.~Gonzalez, and J.~Fernandez-Rossier, Phys. Rev. B {\bf 88}, 245436 (2013).

\bibitem{ponomarenko-vertical} L.~Britnell, R.V.~Gorbachev, R.~Jalil, B.D.~Belle, F.~Schedin, A.~Mishchenko, T.~Georgiou, M.I.~Katsnelson, L.~Eaves, S.V.~Mozorov, N.M.R.~Peres, J.~Leist, A.K.~Geim, K.S.~Novoselov, and L.A.~Ponomarenko, Science {\bf 335}, 947 (2012).

\bibitem{duan-vertical} W.J.~Yu, Z.~Li, H.~Zhou, Y.~Chen, Y.~Wang, Y.~Huang, and X.~Duan, Nat. Mater.  {\bf 12}, 246 (2013).

\bibitem{butler-mgo} W.H.~Butler, X.-G.~Zhang, T.C.~Schulthess, and J.M.~MacLaren, Phys. Rev. B {\bf 63}, 054416 (2001).

\bibitem{sanvito-mgo} I.~Rungger, O.~Mryasov, and S.~Sanvito, Phys. Rev. B {\bf 79}, 094414 (2009).

\bibitem{kawakami-ferro} J.-R.~Chen, P.M.~Odenthal, A.G.~Swartz, G.C.~Floyd, H.~Wen, K.Y.~Luo, and R.K.~Kawakami, Nano Lett. {\bf 13}, 3106 (2013).

\bibitem{dash-ferro} A.~Dankert, L.~Langouche, M.V.~Kamalakar, and S.P.~Dash, ACS Nano {\bf 8}, 476 (2014). 

\bibitem{kelly} V.M.~Karpan et al., Phys. Rev. Lett. {\bf 99}, 176602 (2007).

\bibitem{yazyev} O.V.~Yazyev and A.~Pasquarello, Phys. Rev. B {\bf 80}, 035408 (2009).

\bibitem{sanvito-smeagol1} A. R.~Rocha, V. M.~Garcia-Suarez, S.~Bailey, C.~Lambert, J.~Ferrer, and S.~Sanvito, Nat. Mater.  {\bf 4}, 335 (2005).

\bibitem{sanvito-smeagol2} A. R.~Rocha, V. M.~Garcia-Suarez, S.~Bailey, C.~Lambert, J.~Ferrer, and S.~Sanvito, Phys. Rev. B {\bf 73}, 085414 (2006).

\bibitem{sanvito-smeagol3} I.~Rungger and S.~Sanvito, Phys. Rev. B {\bf 78}, 035407 (2008).

\bibitem{soler-siesta} J.M.~Soler, E.~Artacho, J.D.~Gale, A.~Garcia, J.~Junquera, P.~Ordej\'{o}n, and D.~S\'{a}nchez-Portal, J. Phys.:Condens. Matter {\bf 14}, 2745 (2002).

\bibitem{das-schottky} S.~Das et al., Nano Lett. {\bf 13}, 100 (2013).

\bibitem{tomanek-ti} I.~Popov, G.~Seifert, and D.~Tomanek, Phys. Rev. Lett. {\bf 108}, 156802 (2012).

\bibitem{banerjee-mo} J.~Kang, W.~Liu, and K.~Banerjee, App. Phys. Lett. {\bf 104}, 093106 (2014).

\bibitem{gwbse}D.Y.~Qiu et al., Phys. Rev. Lett. {\bf 111}, 216805 (2013).

\bibitem{sanvito-substrate} K.~Dolui, I.~Rungger, and S.~Sanvito, Phys. Rev. B {\bf 87}, 165402 (2013).

\bibitem{sanvito-mos2ribbon} K.~Dolui, C.D.~Pemmaraju, and S.~Sanvito, ACS Nano {\bf 6}, 4823 (2012).

\bibitem{ciraci-tmd} C.~Ataca, H.~Sahin, and S.~Ciraci, J. Phys. Chem. C {\bf 116}, 8983 (2012). 

\end{thebibliography}
\end{document}